\documentclass[twocolumn,showpacs,superscriptaddress, preprintnumbers , amsmath,amssymb,letterpaper]{revtex4}

\usepackage{dcolumn}
\usepackage{bm}
\usepackage{amsmath}
\usepackage{multirow}
\usepackage{graphicx}
\begin{document}


 \newcommand{\re}{\mathop{\mathrm{Re}}}
 \newcommand{\im}{\mathop{\mathrm{Im}}}
 \newcommand{\D}{\mathop{\mathrm{d}}}
 \newcommand{\I}{\mathop{\mathrm{i}}}
 \newcommand{\E}{\mathop{\mathrm{e}}}
 \newcommand{\unite}[2]{\mbox{$#1\,{\rm #2}$}}
 \newcommand{\myvec}[1]{\mbox{$\overrightarrow{#1}$}}
 \newcommand{\mynor}[1]{\mbox{$\widehat{#1}$}}
 \newcommand{\rmsemit}{\mbox{$\tilde{\varepsilon}$}}
 \newcommand{\mean}[1]{\mbox{$\langle{#1}\rangle$}}

\title{Observation of  Coherently-Enhanced Tunable Narrow-Band Terahertz Transition \\
Radiation from a Relativistic Sub-Picosecond Electron Bunch Train }
\author{P. Piot} \affiliation{Northern Illinois Center for
Accelerator \& Detector Development and Department of Physics,
Northern Illinois University, DeKalb IL 60115,
USA} \affiliation{Accelerator Physics Center, Fermi National
Accelerator Laboratory, Batavia, IL 60510, USA}\author{Y.-E Sun}  \affiliation{Accelerator Physics Center, Fermi National
Accelerator Laboratory, Batavia, IL 60510, USA}
\author{T. J. Maxwell} \affiliation{Accelerator Physics Center, Fermi National
Accelerator Laboratory, Batavia, IL 60510, USA}\affiliation{Northern Illinois Center for
Accelerator \& Detector Development and Department of Physics,
Northern Illinois University, DeKalb IL 60115,
USA}
\author{J. Ruan}  \affiliation{Accelerator Division, Fermi National
Accelerator Laboratory, Batavia, IL 60510, USA}
\author{A. H. Lumpkin}  \affiliation{Accelerator Division, Fermi National Accelerator Laboratory, Batavia, IL 60510, USA}
\author{M. M. Rihaoui} \affiliation{Northern Illinois Center for
Accelerator \& Detector Development and Department of Physics,
Northern Illinois University, DeKalb IL 60115,
USA}
\author{R. Thurman-Keup}  \affiliation{Accelerator Division, Fermi National Accelerator Laboratory, Batavia, IL 60510, USA}

\preprint{PREPRINT FERMILAB-PUB-10-543-APC}
\date{\today}

\begin{abstract}
We experimentally demonstrate the production of narrow-band ($\delta f/f \simeq20$\% at $f\simeq 0.5$~THz)  THz transition radiation with tunable frequency over [0.37, 0.86]~THz.  The radiation is produced as a train of sub-picosecond relativistic electron bunches transits at the vacuum-aluminum interface of an aluminum converter screen. We also show a possible application of modulated beams to extend the dynamical range of a popular bunch length diagnostic technique based on the spectral analysis of coherent radiation.
\end{abstract}
\maketitle
Techniques available so far to produce THz radiation are either limited (optical techniques do not easily provide control of the THz pulse properties~\cite{masayoshi}) or expensive (free-electron lasers require the operation of a large electron accelerator~\cite{biedron}). Motivated by these facts, several groups are exploring alternative ways of producing THz radiation either based on low-energy (keVs) electron beams~\cite{Wachtel,Brau1,Kumar1,prokop} or compact (MeVs) accelerators~\cite{greg,krafft,bosco,neumann}.  In this Letter we demonstrate the use of a 14-MeV pre-modulated electron beam to generate coherently-enhanced transition radiation (TR).  Transition radiation is produced whenever a charged particle crosses the interface of two media with different dielectric constants~\cite{ginzburg}. The TR produced by a single particle is spectrally broadband. The  method we present is general and can be straightforwardly applied to any radiative electromagnetic processes~\cite{gover}.

 \begin{figure}[hhhhh!!!!!!!!!!!!]
\centering
\includegraphics[width=0.46\textwidth]{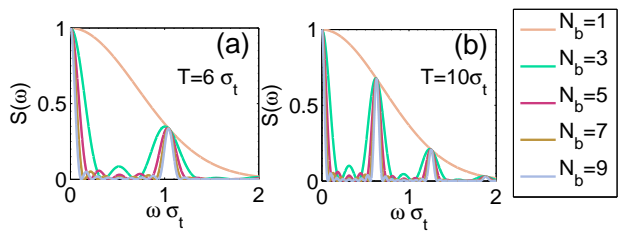}
\caption{(color online) Bunch form factor $S(\omega)$ associated with a train of $N_b$ identical Gaussian bunches separated by $T=6\sigma_t$ (a) and  $T=10\sigma_t$ (b) where $\sigma_t$ is the bunches' rms duration.}
\label{fig:bff}
\end{figure}

In general, the spectral angular fluence emitted by a bunch of $N\gg 1$ electrons from  any electromagnetic process  is related to the single-electron spectral fluence,  $\frac{d^2W}{d\omega d\Omega}\big|_1$,  via $\frac{d^2W}{d\omega d\Omega }\big|_N\simeq \frac{d^2W}{d\Omega d\omega}\big|_1 [N+N^2|S(\omega)|^2]$ where $\omega\equiv 2\pi f$ ($f$ is the frequency) and $S(\omega)$, the bunch form factor (BFF), is the intensity-normalized Fourier transform of the normalized charge distribution $S(t)$~\cite{saxon}. The former equation assumes the bunch can be approximated as a line charge distribution and is practically valid as long as the rms bunch duration $\sigma_t$ and transverse size $\sigma_{\perp}$ satisfy $\sigma_{\perp} \ll c \sigma_t/\gamma$ where $\gamma$ is the Lorentz factor and $c$ the velocity of light. When the BFF approaches unity, $\frac{d^2W}{d\omega d\Omega }\big|_N \propto N^2$ and the radiation is termed as ``coherent radiation". Considering a series of $N_b$ identical bunches  with normalized distribution $\Lambda(t)$  we have $S(t)=N_b^{-1} \sum_{n=1}^{N_b} \Lambda(t+nT)$ (where $T$ is the period between the bunches) giving  $|S(\omega)|^2 =  \xi  |\Lambda(\omega)|^2$ where $\Lambda(\omega)$ is the Fourier transform of $\Lambda(t)$ and the intra-bunch coherence factor  $\xi\equiv N_b^{-2} \sin^2(\omega N_b T/2)/[\sin^2(\omega T/2)]$ describes the enhancement of radiation emission at the resonant frequencies $\omega_n = 2\pi n /T $. The resonant enhancement is characterized by a full-width half-max (FWHM) value  $\delta f \simeq 2 f_1 [ \sqrt{3/(N_b^2-1)}]/\pi$ (for $N_b> 1$). The BFFs computed for bunch train with different parameters are displayed in Fig.~\ref{fig:bff}.

The production of a train of bunches with temporal separation $T\le 1$~ps suitable for producing THz radiation is challenging. Most of the techniques developed and tested so far have severe limitations~\cite{bosco,ychuang,yuelin,neumann}. In the experiment presented in this Letter, we implemented an alternative scheme based on the exchange of phase space coordinates between the transverse and longitudinal degrees of freedom~\cite{emmaprstab,sunPRL}.
 \begin{figure}[hhhhh!!!!!!!!!!!!]
\centering
\includegraphics[width=0.40\textwidth]{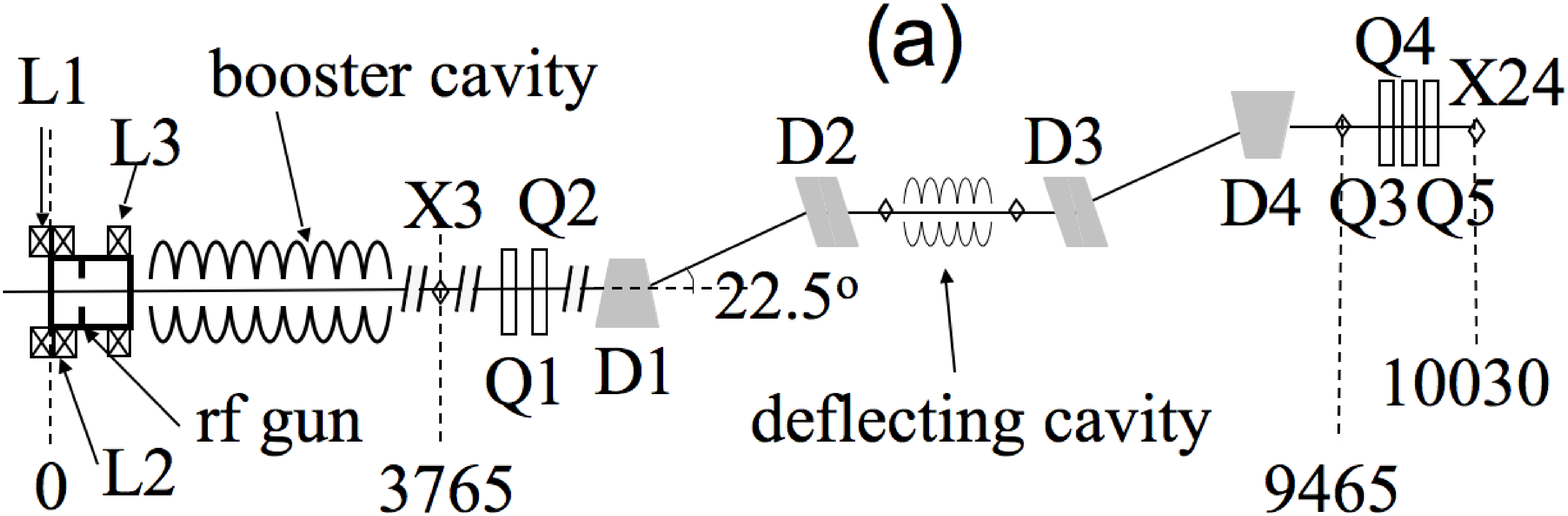}
\includegraphics[width=0.40\textwidth]{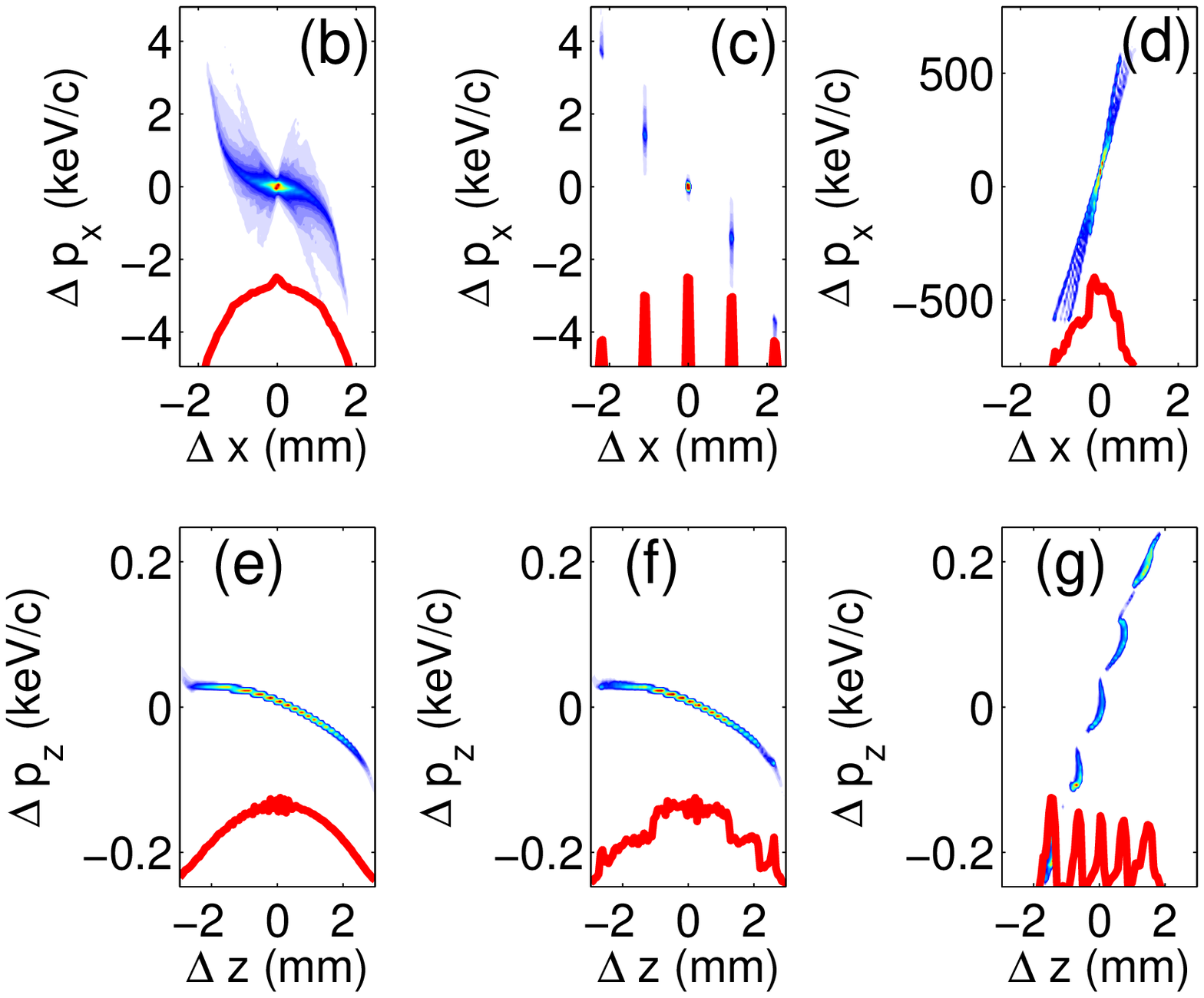}
\caption{(color online) Diagram of the A0 photoinjector with the distances in mm (a). The ``L'', ``Q'' and ``D'' respectively refer to solenoidal lenses, quadrupole and dipole magnets. The density plots are snapshots of the centered transverse $(\Delta{x},\Delta{ p_x})$ (b-d) and longitudinal $(\Delta{z},\Delta{p_z})$ (e-g) phase spaces upstream (b,e) and downstream (c,f) of X3, and at X24 (d,g). The red traces are the associated horizontal (upper row) and longitudinal (bottom row) projections.}
\label{fig:setup}
\end{figure}

\begin{figure}[hhhhh!!!!!!!!!!!!]
\centering
\includegraphics[width=0.40\textwidth]{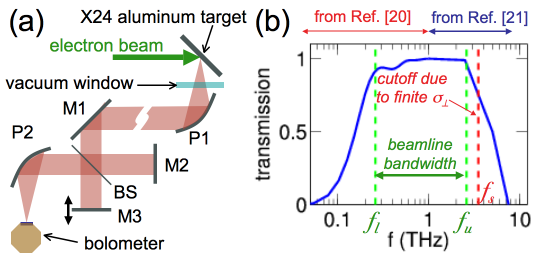} 
\caption{(color online) THz beamline and autocorrelator (a) and corresponding transmission function (b).  P1, P2 are off-axis parabolic mirrors, M1, M2 are planar mirrors,  M3 a movable mirror, and BS is an Inconel\texttrademark-coated beam splitter. \label{fig:beamline}}
\end{figure}

The experiment was performed at the Fermilab's A0 Photoinjector~\cite{carneiro}; see Fig.~\ref{fig:setup}. In brief, electron bunches are generated via photoemission from a cesium telluride photocathode located on the back plate of a 1+1/2 cell radio-frequency (rf) cavity operating operating on the TM$_{010}$ $\pi$-mode at 1.3~GHz (the ``rf gun"). The rf gun is surrounded by three solenoidal lenses (L1, L2, and L3) that control the beam's transverse size and divergence. The beam is then accelerated in a 1.3-GHz superconducting rf cavity (the booster cavity) to $\sim 14$~MeV. Downstream of the booster cavity, the 500-pC bunch is intercepted by a multislit mask consisting of 48-$\mu$m wide slits with 1-mm spacing thereby producing a transversely-segmented beam with total charge of $\sim 15$~pC. The beam is transported, with a set of quadrupole magnets,  to the phase space exchange (PEX) beamline which consists of a liquid-Nitrogen-cooled deflecting cavity operating on the TM$_{110}$-like  $\pi$-mode at 3.9~GHz,  located between two dispersive sections; see Fig.~\ref{fig:setup}. The simulated transverse and longitudinal phase spaces evolution along the accelerator beamline is shown in the sequence of plots (b-g). The simulations, performed with the particle-in-cell package {\sc impact}~\cite{ji},  illustrate how the transversely-segmented beam upstream of the PEX beamline [Fig.~\ref{fig:setup} (c)] is converted into a longitudinally-segmented beam downstream of the PEX beamline [Fig.~\ref{fig:setup} (g)].  Varying the settings of the quadrupole magnets upstream of the PEX beamline [Q1 and Q2 in Fig.~\ref{fig:setup} (a)] enables the control of the final longitudinal phase space parameters and consequently the bunch train parameters. Downstream of the PEX beamline, coherently-enhanced transition radiation (CTR) is generated as the beam transits the vacuum-aluminum interface of an aluminum TR converter labeled as X24 in Fig.~\ref{fig:setup}. The radiation pulse is extracted through a $z$-cut crystal-quartz vacuum window and transported to a Michelson interferometer where the signal is recorded with a Helium-cooled bolometer; see Fig.~\ref{fig:beamline} (a).
The low-frequency suppression and cutoff at $f_l \simeq  0.26$~THz of the beamline's transmission function [see Fig.~\ref{fig:beamline} (b)] is mainly a consequence of the finite size of the X24 aluminum mirror and vacuum window and was simulated using a vector-diffraction model for propagating the TR wave-front~\cite{maxwell}.  At  high frequencies, the measured  transmission for the $z$-cut quartz vacuum window  was used~\cite{zcut} giving a high-frequency cutoff at $f_u \simeq 2.5$~THz.  During our experiments, the beam was focused on  the TR converter with the help of  quadrupole magnets Q3, Q4 and Q5 to $\sigma_{\perp} \simeq 400$~$\mu$m. We therefore expect the beam transverse size to introduce a high-frequency cutoff at $f_s \simeq \gamma c /[2\pi \sigma_{\perp}]\simeq 3.5$~THz beyond the cutoff imposed by the $z$-cut window. The finite bandwidth $\Delta f =[0.26, 2.5]$~THz of the detection system does not affect the measurements reported below but ultimately prevents us from resolving bunch structures below $\sim 20$~$\mu$m (70~fs). 

\begin{figure}[hhhhh!!!!!!!!!!!!]
\centering
\includegraphics[width=0.42\textwidth]{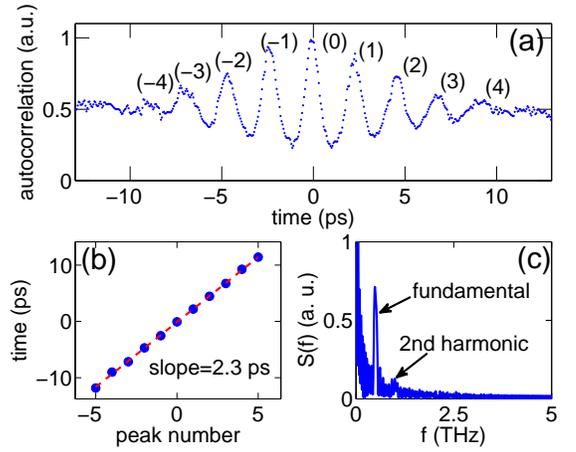}
\caption{(color online) Measured autocorrelation function (a), corresponding peak locations (b) and deduced spectrum for the detected radiation (c). The numbers on plot (a) refer to the peak number. }\label{fig:autocorr}
\end{figure}

For current-modulated beams, the autocorrelation function is multipeaked and the peak separation provides the bunch-to-bunch separation within the bunch train. A Fourier transform of the autocorrelation function directly provides the spectral power density associated with the detected radiation. The spectrum can then be used to infer the resonant frequency and the associated resonance width. An example of measured autocorrelation and computed peak separation are shown in Fig.~\ref{fig:autocorr}.  The bunch spacing is $T=2.3$~ps and the spectrum,  computed using a fast-Fourier transform (FFT) algorithm, exhibits the corresponding fundamental $f_1\simeq 0.43$~THz and second harmonic  $f_2\simeq 0.86$~THz resonant  frequencies.

\begin{figure}[hhhhh!!!!!!!!!!!!]
\centering
\includegraphics[width=0.44\textwidth]{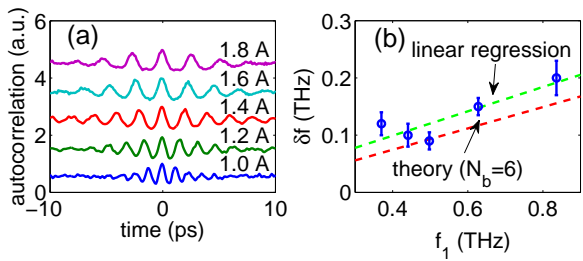}
\caption{(color online) Measured autocorrelation functions (with offset for clarity) for different current settings  (indicated on the plot) of the quadrupole magnet Q1 (a) and corresponding fundamental frequency and FWHM bandwidth (b).}\label{fig:CTRprop}
\end{figure}

A series of autocorrelation functions was recorded for different settings of the upstream quadrupole magnets Q1 and Q2. The measurements taken for Q2=1~A and different values of Q2, are displayed in Fig.~\ref{fig:CTRprop} (a) and illustrate the capability of the method to produce bunch trains with variable spacing thereby providing control over the resonant frequency of the TR; see Fig.~\ref{fig:CTRprop} (b). In our experiment we were able to scan the frequency over $f_1\in [0.37, 0.86]$~THz while maintaining a relative bandwidth below $25 $\%. During this measurement, a train of 6 bunches was produced as inferred by observing the beam's transverse density downstream of X3. The discrepancy between the theoretical model described earlier [red line in Fig.~\ref{fig:CTRprop} (b) with $N_b=6$] and a linear regression of the experimental data (green line) is attributed to the non-uniformity in charge and final duration spread for the bunches across the train. Such non uniformities among the bunches result in a larger FWHM bandwidth.

\begin{figure}[hh!!!!]
\centering
\includegraphics[width=0.44\textwidth]{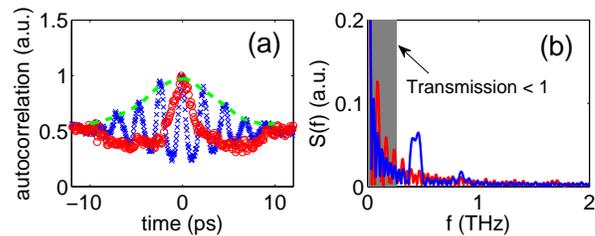}
\caption{(color online) Autocorrelation functions obtained with (blue) and without (red) inserting X3 slits (a) and corresponding bunch form factor (b). The green dashed line in (a) represents the envelop of the multipeaked autocorrelation function (obtained with X3 slits inserted).}\label{fig:CTRdiag}
\end{figure}

Finally, the analysis of the CTR spectrum provides information on the bunch's temporal distribution~\cite{barry}. Widely-used CTR-based diagnostics such as these are limited by the finite bandwidth of the detection system.  Our system cannot measure ``long'' bunch with rms duration larger than $\sim 1/f_l\simeq 3$~ps; see Fig.~\ref{fig:beamline}. Modulating the longitudinal density with a frequency within the bandwidth of the detection system allows for the measurement of these long bunches~\cite{muggli}. 
A direct demonstration of this feature appears in Fig.~\ref{fig:CTRdiag} where the autocorrelation functions of a long  bunch were measured with and without intercepting the beam with the X3 slits. 
For the nominal bunch most of the BFF content is at frequencies below $f_l$ [shown as shaded area in Fig.~\ref{fig:CTRdiag} (b)] thereby resulting in a autocorrelation function with strongly suppressed signal for times $|t| \gtrsim  3$~ps; see Fig.~\ref{fig:CTRdiag} (a).  In the case of the modulated beam the modulation leads to an enhancement of the BFF at $\sim 0.43$~THz so that information on the nominal bunch appears within the detectable bandwidth resulting in a modulated autocorrelation with its envelope representative of the nominal bunch shape. \\

We are indebted to M. Church, E. Harms , E. Lopez, R. Montiel, W. Muranyi, J. Santucci, C. Tan, and B. Tennis for their support. The work was supported by the US DOE Contracts No. DE-FG02-08ER41532  with Northern Illinois University and  No. DE-AC02-07CH11359 with the Fermi Research Alliance, LLC.

\end{document}